\documentclass[10pt,letterpaper]{article}
\usepackage{opex3}
\usepackage{ae}
\usepackage{epsfig}
\usepackage{epsfig}
\usepackage{color}
\usepackage{dcolumn}
\usepackage{graphicx}
\usepackage{graphics}
\usepackage{amsfonts}
\usepackage{amsmath}
\usepackage{amssymb}

\begin{document}

\title{Versatile two-dimensional potentials for ultra-cold atoms}

\author{S. K. Schnelle, E. D. van Ooijen, M. J. Davis, N. R. Heckenberg and H.
Rubinsztein-Dunlop}

\address{School of Physical Sciences, University of Queensland, Brisbane, QLD 4072, Australia.}

\email{sschnell@physics.uq.edu.au}

\begin{abstract}
We propose and investigate a technique for generating smooth
two-dimensional potentials for ultra-cold atoms based on the rapid
scanning of a far-detuned laser beam using a two-dimensional acousto-optical
modulator (AOM).  We demonstrate the implementation of a feed-forward
mechanism for fast and accurate control of the spatial intensity of the laser
beam, resulting in improved homogeneity for the atom trap.  This
technique could be used to generate a smooth toroidal trap that would
be useful for static and dynamic experiments on superfluidity and persistent currents
with ultra-cold atoms.
\end{abstract}


\section{Introduction}

Before the experimental realization of Bose-Einstein condensation
(BEC) \cite{Anderson}, the phenomenon of superfluidity was studied
almost exclusively in the context of superfluid helium
\cite{Kapitza}.  One of the theoretical difficulties with this
system is that liquid helium is not dilute, and so making progress
with microscopic theory is troublesome.  In contrast, interactions
in dilute Bose-condensed gases are weak, which enables accurate
modeling of these quantum many-body systems, both with the
mean-field Gross-Piteavskii equation, and with more advanced quantum
simulation techniques.  Combined with the flexibility of experiments
on ultra-cold atoms,  it seems that dilute gas BECs are potentially
useful systems to further develop our understanding of
superfluidity.

A number of experiments probing the superfluidity of condenstates
have been performed.  Raman \emph{et al.} provided evidence of a critical
velocity by moving a tightly focussed blue-detuned laser through a
trapped condensate and found that above a certain speed there
was significant heating \cite{Raman}. Marago \emph{et al.} \cite{Marago}
observed that a single frequency irrotational scissors mode
excitation existed in a BEC,  as compared with two thermal modes
with different frequencies above the critical temperature.  The
observation of vortices \cite{Haljan} and vortex lattices
\cite{Abo-Shaeer} also provides evidence of the superfluidity of a
BEC.

An important consequence of superfluidity is the phenomenon of
persistent currents, where fluid flow in a multiply connected
potential exhibits no measurable dissipation.  For example, in 1964
Reppy and Depatie were able to set up a flow of superfluid helium in
a torus shaped vessel that showed no reduction in angular momentum
over a 12 hour period \cite{Repy}.  Performing similar experiments
with a Bose-Einstein condensate could be expected to provide useful
information about the superfluid behavior of these systems.

The observation of persistent currents with a BEC
 requires the construction
and loading of a toroidal shaped trap.  A large number of
theoretical papers on BEC in toroidal traps have been written over
the last ten years, and a number of proposals and demonstrations of
ring traps have been described in the literature. These include
magnetic traps \cite{Arnold,Gupta,Sauer}, rf-dressing \cite{Morizot}
and optical methods using spatial light modulators \cite{Padgett}.
Most of these traps tend to be large, which would give increased
sensitivity for e.g. Sagnac interferometry, but are problematic for
the study of superfluidity.  In particular it is difficult to ensure
homogeneity of the potential over a large area, or form a
multiply-connected condensate delocalized over the entire ring
structure. Significantly, Ryu \emph{et al.} \cite{Ryu} recently
reported the first observation of a persistent current in a
multiply-connected BEC in a magnetic trap plugged with an optical
dipole potential.  However, the lifetime of the superflow was
limited to 10 s due to a drift of the magnetic trap relative to the
dipole potential, and the trap depth varied about the the torus
(although the variation was less than the chemical potential of the
BEC.) For a careful study of superfluidity it is desirable for the
toroidal trap to have a constant potential depth about the entire
ring, with fluctuations much smaller than the chemical potential.

By spatially scanning a focussed laser beam at a frequency much
higher than the atoms can respond to,  a range of time-averaged 2D
potential geometries can be created.  This procedure is similar of
the original Time Orbiting Potential (TOP) trap developed by Petrich
\emph{et al.} in the original quest for BEC \cite{Petrich}.  A
number of time-averaged traps have been realized in the past, in
particular \cite{Rudy,Friedman,Milner} who used blue-detuned beams
for trapping thermal atoms and \cite{Ahmadi} who scanned their
CO$_2$ to increase the volume of capture in the transfer of their
atoms from their MOT to dipole trap.   Recently, Lesanovsky \emph
{et al.} \cite{Lesanovsky} proposed a time averaging procedure for
adiabatic rf potentials and magnetic fields to generate a smooth 3D
ring trap for atoms; however this is yet to be realized in practise.

In this work we propose a versatile, all-optical trap based on
a fast-scanning laser beam using a two-dimensional acousto-optical
modulator (AOM). We focus in particular on generating a smooth
toroidal trap, and we show that that using a feed-forward system we can
produce a relatively small and homogeneous ring potential suitable
for experiments concerning superfluidity. Moreover, we show that
this system is a versatile tool for producing basically any 2D
static or dynamical potential. Trapping in the third dimension
perpendicular to the ring can be provided, for example, by a
light sheet provided by a second laser, resulting in a
fully three-dimensional trap for ultra-cold atoms.

\section{Theory of a ring trap}

\begin{figure}[htbp]
  \centering
    \includegraphics[width=78mm]{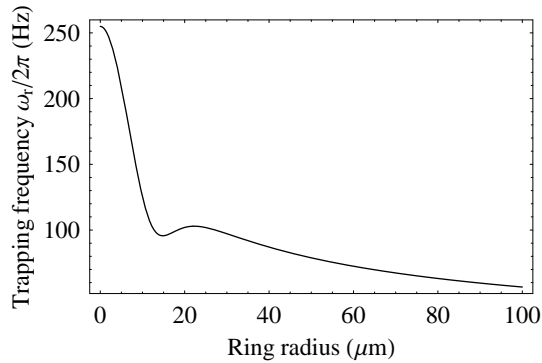}
  \caption{Trapping frequencies as a function of ring size for a beam waist of $25 \mu $m.
  The minimum in this plot shows the regime where the radius $a$ of the ring is of the same size as the focus of the laser focus and the minimum of
  the potential in the center of the trap $< 0$. }
  \label{freq}
\end{figure}

Ultra-cold atoms may be confined in a conservative all-optical potential
using a
focused laser beam that is sufficiently far detuned from the atomic transition
so as to prevent spontaneous emission.
For a static laser beam with spatial intensity $I(\vec{r})$ and detuning $\Delta = \omega_L
- \omega_A$ from the optical transition frequency $\omega_A$, the
optical dipole potential is given by

\begin{equation}
U(\vec{r}) \approx \frac{3 \pi c^2}{2 \omega_A^3}
\frac{\Gamma}{\Delta} I(\vec{r})
\end{equation}

\noindent for large detunings \cite{Allen}. When such a  beam is
scanned sufficiently fast in a periodic manner, the dipole potential
should be time-averaged over one cycle.  For a beam with a two
dimensional intensity profile $I(x,y) = \frac{2P}{\pi w^2} \exp{
\left(\frac{-2(x^2 + y^2)}{w^2} \right)}$ scanned in a ring with
radius $a$, the resulting radial dependent average intensity is
\begin{equation}
 I(r)=\frac{4 P}{w^2} \exp \left( \frac{-2 (r^2+a^2)}{w^2} \right) \mathcal{I}_0
 (\zeta),
\end{equation}

\noindent where $\mathcal{I}_0 (\zeta)$ denotes the modified zeroth-order
Bessel function with $\zeta = 4 a r/w^2$ and
$w$ the beam waist of the focused beam trap.  The  trap minimum occurs at $r=a$, and has a radial trapping frequency
of

\begin{equation}
 \omega_r = \sqrt{\frac{2 \kappa}{m} P \exp (-\frac{4 a^2}{w^2}) \left[ \frac{64 a^2}{w^6} (\mathcal{I}_0(\zeta)-\mathcal{I}_1(\zeta)) - \frac{8}{w^4} (\mathcal{I}_0(\zeta)-\mathcal{I}_1(\zeta)) \right]
 },
\end{equation}

\noindent with $\mathcal{I}_n$ the modified $n$-th order Bessel
functions and $\kappa = -\frac{3 \pi c^2}{2 \omega_A^3}
\frac{\Gamma}{\Delta}$.  We find that $\omega_r \propto 1/\sqrt{a}$ for $a
> w$ (Fig. \ref{freq}).

\section{Experimental setup and results}

To create a ring potential an intensity locked, red-detuned laser
beam is rapidly scanned in space using a
two-dimensional AOM (IntraAction, DTD-274HD6), with two independent AOMs
mounted perpendicularly with respect to the diffraction directions.
The deflection angle of a beam in an AOM depends only on the
frequency of the traveling radio frequency (rf) wave inside the
crystal. Therefore, by modulating the rf frequencies that are fed
into the crystals the beam can be scanned in arbitrary 2D patterns.
These scanning frequencies are high compared to the trapping
frequencies preventing the atoms from following the beam movement.
Rapid scanning frequencies also prevent parametric heating that
takes place close to the trapping frequencies and multiples thereof
\cite{Savard}. When changing the angle of deflection the diffraction
efficiency of the AOM and therefore the intensity of the deflected
beam varies. To create homogenous intensity distributions this
variation needs to be compensated by controlling the rf power
fed into the AOM. This is possible as the diffraction efficiency of
an AOM is linear to the rf power driving it.

\begin{figure}
\centering
  \includegraphics[width=0.5\textwidth,angle=-90]{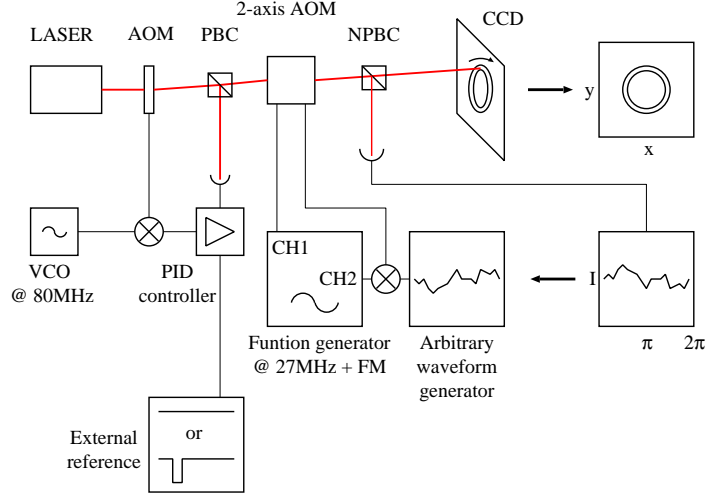}
  \caption{Feed-forward technique to cancel the effect of different diffraction efficiencies
depending on the angle of deflection. When the intensity locked
laser beam is scanned it is imaged on a CCD camera and some part of
it is deflected onto a photo diode. The photo diode gives an
intensity profile  that is then inverted and mixed with the rf
signal that drives the AOM. Extra modulations of the intensity over
the ring can be accomplished with an additional signal to the
intensity lock.}
  \label{setup}
\end{figure}

As the variation in intensity with deflection angle is
deterministic, a feed forward technique is used to compensate for
the intensity variation. A small amount of the trapping beam is
split onto a photodiode to measure the power to control the light
distribution, while the remainder of the beam is imaged on a CCD
camera. The diameter of the photodiode is 800 $\mathrm{\mu}$m, much
bigger than the ring diameter. The photodiode is fast enough to
resolve single scans of the beam thus creating an intensity profile
of the ring. To compensate for the changes in diffraction efficiency
the inverse of this intensity profile is then mixed with the rf
signal driving one of the two AOMs and fed into this AOM (see Fig.
\ref{setup}).

Using this setup stable toroidal traps are created. For these traps
the laser beam is focused down to a spot size of 25 $\mu$m and then
scanned to create a ring radius of 44 $\mu$m. The scanning speed is
51 kHz, well above any of the resulting trapping frequencies. Using
the feed forward technique described above, the intensity
fluctuations about the ring are less than $1.5$\%. Here, we observed
no correlation between the noise and the initial intensity
fluctuations or the frequency modulation of the rf. We therefore
conclude that the remaining noise level is due to electronic noise.
Further analysis shows that for our setup the noise was limited by
the noise of the photo diode. In principle, the limit in intensity
stability of the this method is only restricted to the initial laser
fluctuation and the accuracy of the rf synthesizer and electronics,
which is better than typically $10^{-3}$. The 2D AOM operates at
relatively low efficiency/rf-power, where the deflected laser power
is linearly proportional with the rf-power. It should be noted that
using a feedback system would be preferential, but electronically
far more complicated, as this would require a relatively high
frequency feedback system.

To study superfluidity with a BEC, it is desirable that the bottom
of the ring potential should be as smooth as possible. Therefore,
the spatial intensity distribution throughout the trap should be as
homogenous as possible, and in particular fluctuations in trap depth
should be much smaller than the chemical potential of the
condensate.  In the Thomas-Fermi approximation for a condensate in a
ring trap with a trapping frequency of $\omega_z$ in the $z$
direction we find
\begin{equation}
 \mu = \hbar \sqrt{\, \omega_r \,
\omega_z}\sqrt{\frac{3 N a_s}{4 a}},
\end{equation}
where  $g = 4 \pi \hbar^2 a_s/m$, $m$ is the atomic mass and $a_s$
is the the s-wave scattering length.

In the case of a trap with beam waist $w $= 25 $\mu$m, ring radius
$a = 44$ $ \mu$m, power $P = 5$ mW and at a wavelength of $1064$ nm,
this gives a trap depth of $479$ nK, and a corresponding variation
in trap depth of $7.2$ nK. The trapping frequency in the radial
direction of such a trap is $\omega_r = 2 \pi \cdot 83$ Hz and
assuming a trapping frequency in $z$-direction of $\omega_z = 2 \pi
\cdot 350$ Hz and $2 \cdot 10^{5}$ atoms in the BEC one gets a
chemical potential of $\mu = 36$ nK. Comparing the chemical
potential to the fluctuation of the trap depth one can easily see
that this trap is a feasible tool for the investigation of
superfluidity in degenerate gases.

In order to load a BEC into such a ring potential, we propose to
start with a cloud of cold atoms created in a crossed dipole trap,
consisting of a focussed laser beam and a sheet of light in the
horizontal plane with respect to gravity, crossing the focus of the
first laser beam. Atoms can be evaporated in such a trap close to or
beyond $T_c$, after which the focus of the first beam can be rotated
as described above with increasing amplitude and further evaporation
can be applied. Alternatively, the cold atoms from the all-optical,
static trap can be loaded instantaneously into an overlapping ring
trap, as proposed by Wright \emph{et al.} \cite{wright}.

\begin{figure}[htbp]
  \centering
    \includegraphics[width=78mm]{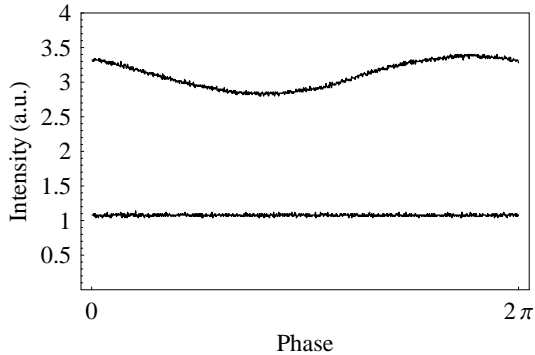}
  \caption{Intensity profiles of the rings without (top curve) and with (bottom curve) feed-forward. Corresponding CCD images
  are shown in figure \ref{dynamic}, where the first image shows the uncorrected and the second image the corrected ring.}
  \label{Labelname}
\end{figure}

\section{Time-averaged optical potential for a BEC}

As described above, for the ring trap the rotation frequency of the
beam has to be much higher than the trapping frequencies for the
potential to be time-averaged and to prevent parametric heating.
Consequently, an atom in this ring potential experiences an optical
trap with a duty cycle of typically 20{\%}.

To test the time-averaging effect of such a trap the following
experiment was performed. A thermal cloud close to the critical
temperature $T_{c}$ was created in a cigar-shaped magnetic trap,
with trapping frequencies of $160 \times 6.8$ Hz, on a chip
\cite{Chris} after which the trap was overlapped with a dipole trap
with a beam waist of 11 $\mu$m crossing the middle of the long axis
of the trap. For a dipole trap with a depth of 250 $\mu$K this
results in the formation of a BEC with a 15.3{\%} BEC fraction in
equilibrium due to the change in the density of states
\cite{Stamper-Kurn1998a}.
\begin{figure}
 \centering
 \includegraphics[width=78mm]{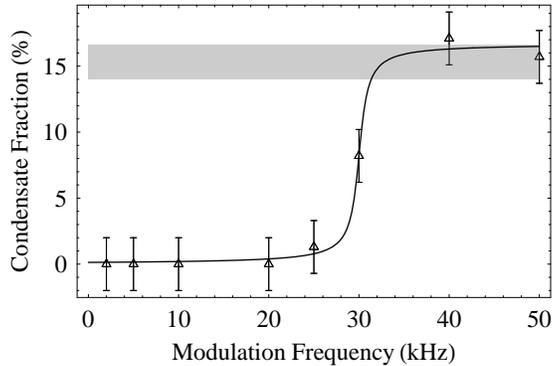}
 \caption{Condensate fraction of a BEC in a magnetic trap with an optical dipole trap superimposed run with a 20\% on/off duty cycle. Initial condensate fraction without dipole trap was 15.3\% and is indicated in gray. The line through the data points acts only as a guide to the eye.}
\label{modul}
\end{figure}
Subsequently, the same experiment was performed where the intensity
of the laser was increased by a factor of five but modulated with a
duty cycle of 20{\%} to ensure the same time-averaged trap depth.
Figure \ref{modul} shows the condensate fraction as a function of
the modulation frequency, showing that above a frequency of 30 kHz
the condensate fraction approaches the same value as for a DC
optical trap (indicated with gray). The rotation frequencies in the
presented ring trap is 51 kHz, with a maximum possible frequency of
100 kHz, and is therefore sufficiently high for the trap to be
time-averaged when ultra-cold atoms close to $T_{c}$ are considered.

\section{Arbitrary and dynamic 2D potentials}

\begin{figure}
  \centering
  \includegraphics[width=0.75\textwidth]{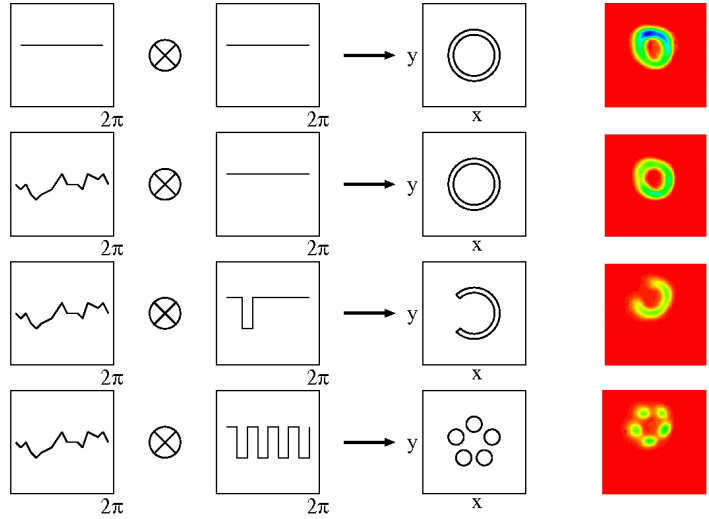}
  \caption{Different possible waveforms for the function generator controlling the feed-forward technique
  and the function generator controlling the PID circuit and the resulting patterns. The top pattern shows a
  uniform distribution for both function generators resulting in a ring with non-uniform light distribution. For all the
  other cases the first generator is used to cancel the effects of different diffraction efficiencies and the second generator
  to create patterns of laser power leading to a uniform ring, horse shoe and ring lattice, respectively. The size of the images is 330 x 330$\mathrm{\mu}$m.}
  \label{dynamic}
\end{figure}

As well as the correction signal, it is possible to add an extra modulation
signal that is also synchronized with the rf-modulation frequency
$f_{\mathrm{s}}$.  This results in a static and arbitrary intensity
modulation on top of the 2D potential. For this we use the reference
input of the intensity lock of the setup (PID circuit, see Fig.
\ref{setup}). To show the feasibility of this method to produce
various other 2D trapping geometries, a horse shoe shape and a ring
lattice with five lattice sites have been created and stabilized
(Fig. \ref{dynamic}).

To create 2D potentials that vary with time we can, for example,
introduce a small difference in the frequency of the rf modulation
and the frequency of the extra intensity modulation $f_{\mathrm{p}}$
in such a way that the pattern rotates with the difference of the
two frequencies, $f_{\mathrm{d}}=|f_{\mathrm{s}}-f_{\mathrm{p}}|$.
In this manner, for example,  a ring lattice can be slowly created with
excellent control over all trapping parameters and could be used to
initiate persistent currents as suggested in \cite{Padgett}. Another
method of introducing angular momentum to the system would consist in a
dip in laser power that is slowly moved around the ring to stir the
BEC, similar to the technique used to stir a BEC with a blue
detuned laser beam (e.g. \cite{Abo-Shaeer}).

In a similar fashion we can arbitrarily adjust the shape and the depth
of the potential as a function of time, where the spacial resolution
is limited by the waist of the laser focus. The temporal resolution
is merely limited by the ratio between size of the laser beam in the
AOM-crystal and the sound of speed in the crystal or by the
electronics used, which for both is better than 100
${\mathrm{\mu}}$s.

\section{Conclusion}

We have made use of a feed-forward technique to generate a new,
versatile all-optical ring trap that should be suitable for the
investigation of superfluidity and persistent currents in
Bose-Einstein condensates. It  could also be a  suitable platform
for interferometry and Josephson tunneling experiments
\cite{Banderson}. This could lead to potential applications in the
field of sensitive rotation sensors and SQUID-like devices. Our
system enables us to arbitrarily control the depth and shape of the
potential in an accurate and controlled manner.  It could therefore
be used to produce other 2D trapping geometries and is a useful tool
for both static and dynamic experiments with degenerate gases.

The authors would like to acknowledge the financial support from the
Australian Research Council (ARC).

\end{document}